\documentclass[submission, Phys]{SciPost}
\usepackage{graphicx}
\graphicspath{ {./images/} }

\begin{document}

\begin{center}{\Large \textbf{
Non-unitary Quantum Electronics: \\
Novel Functions from the Edge of the Quantum World}
}\end{center}

\begin{center}
J. Mannhart*,
H. Boschker,
P. Bredol
\end{center}

\begin{center}
Max Planck Institute for Solid State Research, 70569 Stuttgart, Germany\\

* office-mannhart@fkf.mpg.de
\end{center}

\begin{center}
\today
\end{center}


\section*{Abstract}
{\bf
Novel categories of electronic devices and quantum materials are obtained by pipelining the unitary evolution of electron quantum states as described by Schr\"odinger's equation with non-unitary processes that interrupt the coherent propagation of electrons. These devices and materials reside in the fascinating transition regime between quantum mechanics and classical physics.

The devices are designed such that a nonreciprocal unitary state evolution is achieved by means of a broken inversion symmetry, for example as induced at material interfaces. This coherent state evolution is interrupted by individual inelastic scattering events caused by defects coupled to an environment.

Two-terminal non-unitary quantum devices, for example, feature nonreciprocal conductance in linear response. Thus, they are exemptions to Onsager's reciprocal relation, and they challenge the second law of thermodynamics. 

Implementing the device function into the unit cells of materials or meta-materials yields novel functionalities in 2D and 3D materials, at interfaces, and in heterostructures. 
}

\vspace{10pt}
\noindent\rule{\textwidth}{1pt}
\tableofcontents\thispagestyle{fancy}
\noindent\rule{\textwidth}{1pt}
\vspace{10pt}

\section*{Preface}
\label{sec:preface}

This short contribution provides a summary of the key points of the lecture ``Non-unitary Quantum Electronics---A New World beyond Onsager and Clausius" given on 18 July 2020 at the Les Houches Summer School on Emergent Electronic States Confined at Interfaces, which is available at \cite{youtube}. The novel concept of non-unitary quantum electronics arose from the study of electronic states that emerge at interfaces characterized by Rashba coupling  \cite{Mannhart2018}. It turned out that the basic principles behind non-unitary quantum electronics are far more fundamental and general than being constrained to interfaces. To show their importance, we present here the general principles that have emerged at interfaces without mentioning interfaces further, referring the interested reader instead to~\cite{Mannhart2018}.

\section{Introduction}
\label{sec:intro}

Our macroscopic world is rooted in the quantum world. The transition regime between both worlds, in which characteristic length scales are large enough that quantum states lose their phase memory so that non-unitary processes affect the evolution of the quantum states, is home to a palette of unique phenomena. These phenomena are forbidden in the coherent quantum world as described by Schr\"odinger's equation and are also disallowed in the classical regime. Hence, they can be used to realize devices with functions and to design materials with properties that are impossible to achieve by any other means \cite{Bredol2019}. In this presentation, we explain the principles of such non-unitary electronic devices and materials, starting by briefly summarizing the underlying axioms of quantum physics.

The time evolution of a quantum state $| \psi(t) \rangle$ can---at least in principle---be calculated with superb accuracy. For these calculations, two different procedures must be used. These are considered to be two fundamental axioms of quantum physics (see Fig.~1):

(I)~For an isolated quantum system, Schr\"odinger's equation $i \hbar \, d/dt \, | \psi (t) \rangle = H \, |\psi (t) \rangle$ describes the evolution of the state, where $H$ is the Hamiltonian of the system. 

Schr\"odinger's equation given in (I) provides for a unitary time evolution $| \psi (t) \rangle = e^{-iHt/ \hbar} \, | \psi(0) \rangle $, which is thus called because the evolution operator $ U = e^{-iHt/ \hbar }$ is unitary (i.e., $U^\dagger \, U=1$, where $U^\dagger$ is the adjoint of $U$). Like classical physics, the corresponding \textbf{unitary quantum physics} obeys time-reversal symmetry and is deterministic. Note that the essence of quantum information technology consists of evolving a quantum state only by employing a sequence of unitary transformations.

(II) When the quantum system is coupled to many modes or to a large number of degrees of freedom as provided by an environment, a second calculational method different from (I) must be used. Such a coupling provides the possibility to measure an observable $A$. A measurement of $A$ changes the wave function $ | \psi(t) \rangle = \Sigma_n \, c_n (t) \, | \psi_n (t)\rangle$, written here as sum of the eigenfunctions $| \psi_n \rangle$ of $A$ with the complex weight factors $c_n$, to yield $| \psi_i (t)\rangle$ if the measurement result is the eigenvalue $a_i$ of $A$ (von~Neumann projection \cite{vonNeumann}). This transition occurs with a probability of $|c_i |^2$ (Born's rule~\cite{Born1926}).

The transition described by axiom (II) is a \textbf{non-unitary time evolution}. Here, nature makes choices and also breaks time-reversal symmetry. This becomes apparent, for example, by considering that the projection of $| \psi \rangle = \Sigma \, c_n | \psi_n \rangle$ on $| \psi_i \rangle$ entails the loss of information on the prior coefficients $c_n$. What a ``measurement" actually is, continues to be debated since the early days of quantum mechanics (for an overview, see, e.g.~\cite{Wheeler}). For our case, it is only important that inelastic scattering events in which momentum and energy are transferred between an electron and a defect coupled to a thermal bath given, for example, by a macroscopic substrate, are successfully described as non-unitary processes. Trapping of an electron at a defect allows the location of the electron to be determined by measuring the charge of the defect with a field effect transistor, for example.

\begin{figure}[t]
\includegraphics[width=1.0\textwidth]{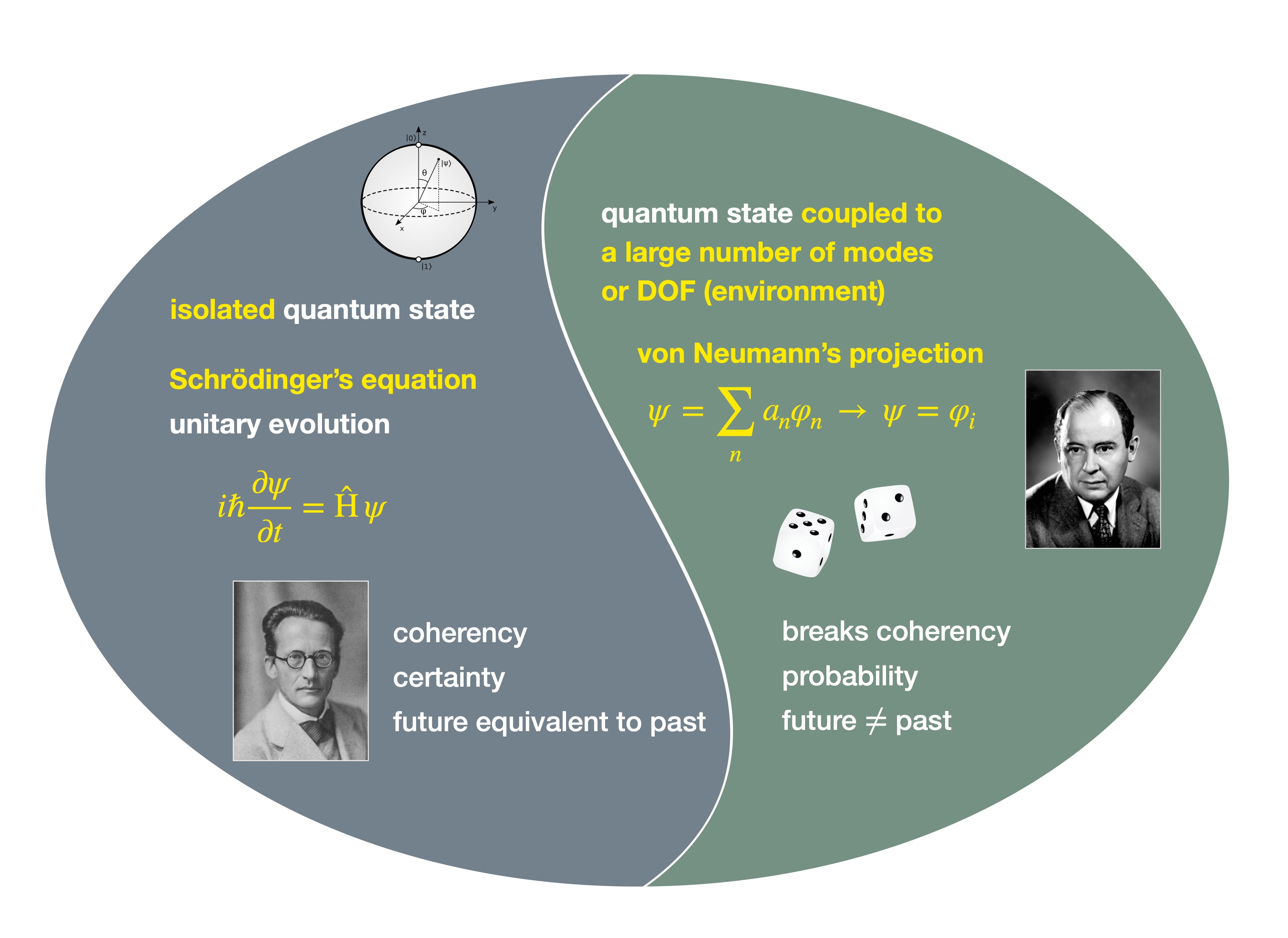}
\caption[width=7cm] {Illustration of the evolution of quantum states according to the calculational processes given by the axioms of quantum physics. If isolated from the environment, a quantum state evolves unitarily as described by Schr\"odinger's equation. If coupled to the environment, the state evolution is accurately described mathematically by Born's rule and von Neumann's projection.  Sources of photos  E.~Schr\"odinger:~ \cite{photoschroe}, J. von Neumann:~\cite{photoneu}.}
\centering
\end{figure}

Several interpretations of quantum mechanics propose that the non-unitary time evolution and the projection of quantum states are simply offspring of the unitary evolution. Of course, these interpretations have to include additional processes to break time-reversal symmetry. In the decoherence interpretation \cite{Joos2013}, for example, time-reversal symmetry is broken by incorporating the procedure of tracing-out the density matrix over the environmental degrees of freedom. This tracing erases information---there is no inverse operation, no ``tracing-in". In the many-worlds interpretation \cite{Everett1957}, it is the splitting of the worlds; a merging of worlds is not foreseen. Independent of the interpretation, innumerable studies have found that calculation procedures (I) and (II) yield an accurate quantitative description of quantum phenomena.

In many cases it suffices to focus only on the unitary evolution of a quantum state. In condensed matter physics, for example, the standard approach to determine the electronic properties of a material consists of establishing the system’s Hamiltonian, for which  Schr\"odinger's equation is then solved. Moreover, quantum information processing builds exclusively on the unitary evolution. Processes (II) are brought into play only to read out the results by measurements performed on $| \psi (t_\textrm{f}) \rangle$, where $t_\textrm{f}$ is the final time of the quantum information processing.

Here, we instead consider quantum systems that utilize both processes (I) and (II). Their evolutions exceed those obtainable by using only process (I). When processes (I) and (II) are combined, novel and unique phenomena emerge in such quantum systems. These exist in the transition regime between the quantum world and the classical world. We will introduce this transition regime next.

\section{The Quantum World, the Transition Regime, and the Classical World}

\subsection{Unitary quantum regime}

As described above, quantum states in the unitary quantum regime evolve \textbf{deterministically} as described by Schr\"odinger's equation. Transport is based on interfering quantum waves, which typically are plane waves representing the energy eigenstates of the system or wave packets built from them. The unitary quantum regime is characterized by \textbf{time-reversal symmetry}; the future is qualitatively equivalent to the past and determined with certainty. The quantum waves are coherent with an infinite decoherence time.

\subsection{Transition regime}

In the transition regime, decoherence lengths and times are comparable to the characteristic scales of the quantum systems. Quantum systems operate in the transition regime if their size or operating temperature causes the inelastic scattering length to be comparable to the system size.

To introduce the dynamics in the transition regime, we consider a quantum ring containing a defect that, with a given probability, acts as a trapping site and therefore as the inelastic scattering center for electrons (Fig.~2). In this case, the characteristic device size is comparable to the inelastic mean free path $l \approx l_\textrm{in}$, and the decoherence time is comparable to the average transit time an electron needs to pass the device. An electron wave packet moves unitarily through the ring as described by Schr\"odinger's equation (I) until the electron is trapped by the defect. The electron momentum is hereby taken up by phonons, which disappear into the thermal bath. This trapping is described by von~Neumann’s projection postulate and Born's rule (II). After a characteristic time, phonon fluctuations cause the electron to leave the trapping center again and to continue propagating on a trajectory  determined by Schr\"odinger's equation (I). This trajectory differs from the one the electron would have followed if the trapping had not occurred, the one that process (I) would have yielded. As the inelastic scattering is not time-reversal symmetric, the overall dynamics of the quantum states is \textbf{not time-reversal symmetric} in the transition regime either, although it includes unitary processes (I). As Born's rule describes the probability of inelastic scattering, the dynamics is also \textbf{stochastic}. In quantum optics, the quantum trajectories method \cite{Carmichael1992,Molmer1993} is a common equivalent method to describe photon dynamics in the transition regime.

\begin{figure}[t]
\centering
\includegraphics[width=0.7\textwidth]{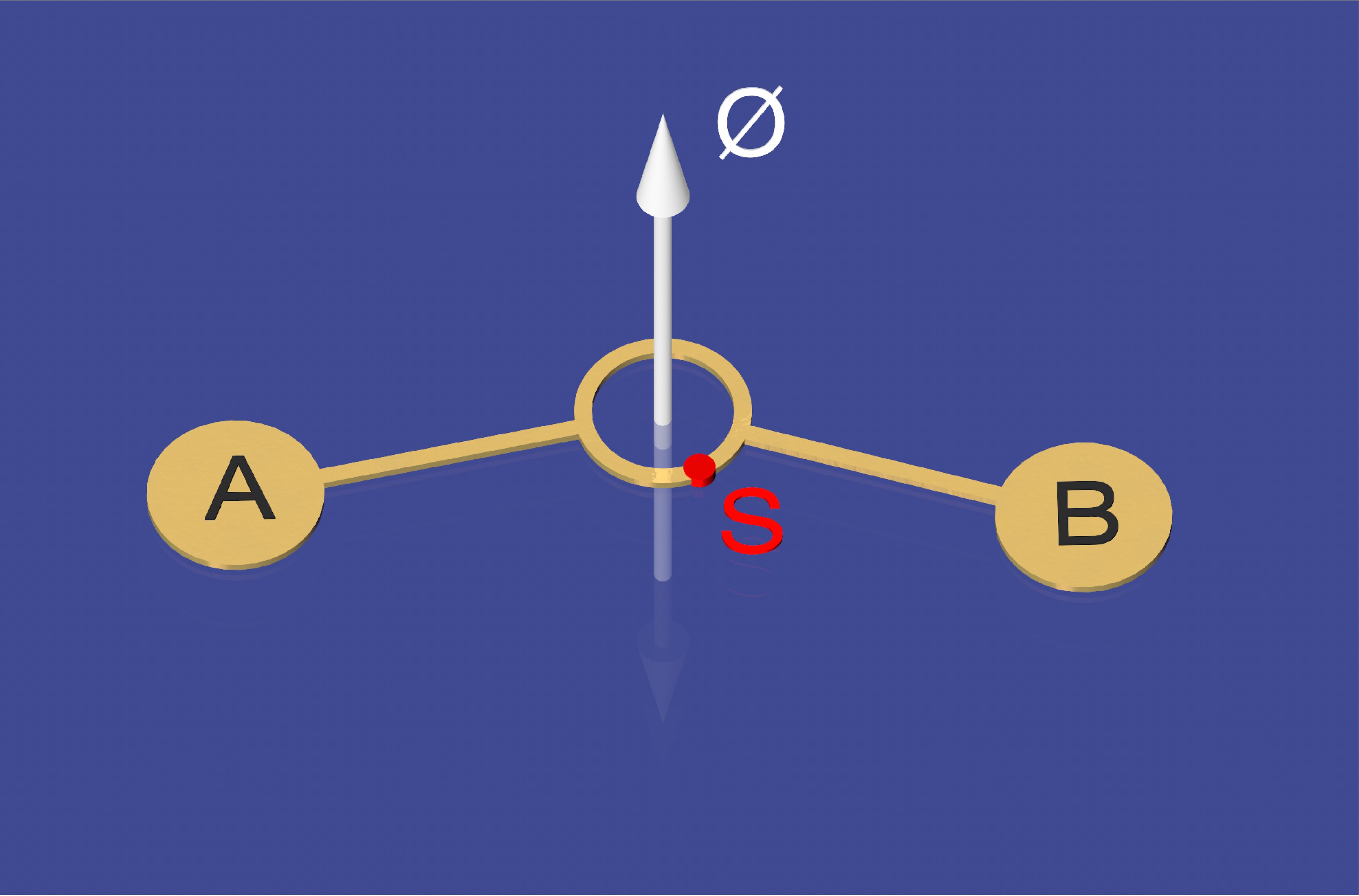}
\caption[width=0.7] {Sketch of an asymmetric quantum ring connected to terminals A and B and penetrated by a magnetic flux $\Phi$. The ring contains a trap site S that acts as the inelastic scattering center. This trap site is coupled via the substrate to a thermal bath. In the configuration considered here, the trap induces an inelastic scattering length or phase-breaking length of the order of the ring circumference $l_\textrm{in} \approx 2 \pi R$. From~\cite{Mannhart2020}.}
\centering
\end{figure}

\vspace{2cm}

The Kubo formalism and its derivations provide a linear-response description of electron dynamics in the transition regime \cite{Kuboformalism}. However, these models disregard the disturbance of the state evolution by non-unitary processes (II). They therefore cannot describe  time-reversal-symmetry-breaking state evolutions induced by processes of type (II) \cite{Bredol2019}. Furthermore, adding decoherence  phenomenologically, the Keldysh formalism \cite{Keldysh1965} has been used to describe nonequilibrium transport in the transition regime (see, e.g., \cite{Pelzer2014}). In addition to these models, more phenomenological ones \cite{Gefen1984,Stern1990,Blanter2000,Hansen2001,Imry2018} are also used to describe the transition regime. Those models blend characteristic properties of time-reversal symmetric classical transport, such as resistance and noise, with unitary quantum transport. They are therefore time-reversal symmetric, too, and must fail to describe phenomena characterized by time-reversal symmetry breaking. 

\subsection{Classical world}

The foundations of classical mechanics are perfectly well captured by Newton's laws or the Hamiltonian formalism. All microscopic mechanical laws are \textbf{deterministic} and \textbf{time-reversal symmetric}. Note that the apparent breaking of time-reversal symmetry by the second law of thermodynamics is caused only by the choice of boundary conditions for a system's temporal evolutions (e.g.,~the universe evolving from a low-entropy state created by the Big~Bang).

Classical or semiclassical transport of electrons can be described by the Drude--Som\-mer\-feld model, which, for example, leads to Ohm's and Kirchhoff's laws. Compared to the transition regime, the scattering events are so numerous and the inelastic scattering length and decoherence time  are so short ($l \gg l_{\textrm{in}}$) that the electron propagation is well described by classical trajectories. Coherent propagation of waves and the interference of quantum states are negligible. Particles, not waves, are transported, and time-reversal symmetry characterizes their dynamics.

\section{The Big Question}

The classical world and the unitary quantum world are well described by a set of fundamental physical laws that are based on the world's symmetries. The derivation of Onsager's powerful and, as found, universally valid reciprocal relation \cite{Onsager}, which describes how a system close to thermodynamic equilibrium relaxes to its ground state, for example, requires time-reversal symmetry of all microscopic processes. Fundamental laws that are valid in the unitary quantum regime and in classical physics are sometimes taken to be universally valid  because classical physics and the unitary quantum world are enormously large domains. These laws therefore tend to be applied also to the transition regime, following the argument that the transition regime represents a mixture of the two adjoining worlds. However, this reasoning is far from clear and deserves to be questioned. Exactly which of the laws based on symmetries or other properties that do not exist in the transition regime are valid in the transition regime? For example, does Onsager's relation, which is based on time-reversal invariance, also hold at the edge of the quantum world? What about the second law of thermodynamics? To answer these questions, we consider a concrete example in which the evolution of quantum states is determined by a combination of non-unitary and unitary processes.

\section{Non-unitary Quantum Ring}

\subsection{Non-unitary evolution}

Let us start by describing the non-unitary process (II) that, in our chosen system, is an essential part of the state evolution. Capturing an electron at a trapping site anchored to a heat bath and its subsequent release yields this process (see Fig.~3).

Consider a single electron moving in the conduction band of a semiconductor such as silicon. The electron is described by a wave packet, which for simplicity is supposed to propagate only in the $\pm x$~direction. We start by having the electron approach a defect, say a cadmium atom, that forms a deep state in the silicon band gap. The arrangement is placed on a macroscopically large substrate; the entire setup is held at a finite temperature~$T$.

When the electron reaches the cadmium, it is captured by the atom with some probability. If the electron is captured, the wave function describing the original electron wave packet is projected onto a defect state as described by (II). The original phase of the electron is lost, and its momentum and energy are coupled into the phonon system of the substrate, which acts as a heat bath. After a random waiting time determined by the Boltzmann distribution, a thermal fluctuation of the phonon system excites the electron back into the conduction band, where the electron again forms a wave packet. As quantified by Born's rule or the related Fermi’s golden rule, the electron then moves either in the $+x$ direction, in the $-x$ direction, or coherently in both. As the spatial symmetry of the setup is unbroken, the probabilities to find the electron later to the left or right of the trap are identical. Such electron behavior is well known. A related configuration is used, for example, in the flash memories of USB sticks to store information, encoded as the presence or absence of electrons in trap sites embedded in the gate stacks of field-effect transistors. In this case, an electron may be trapped for years.

\begin{figure}[t]
\centering
\includegraphics[width=0.5\textwidth]{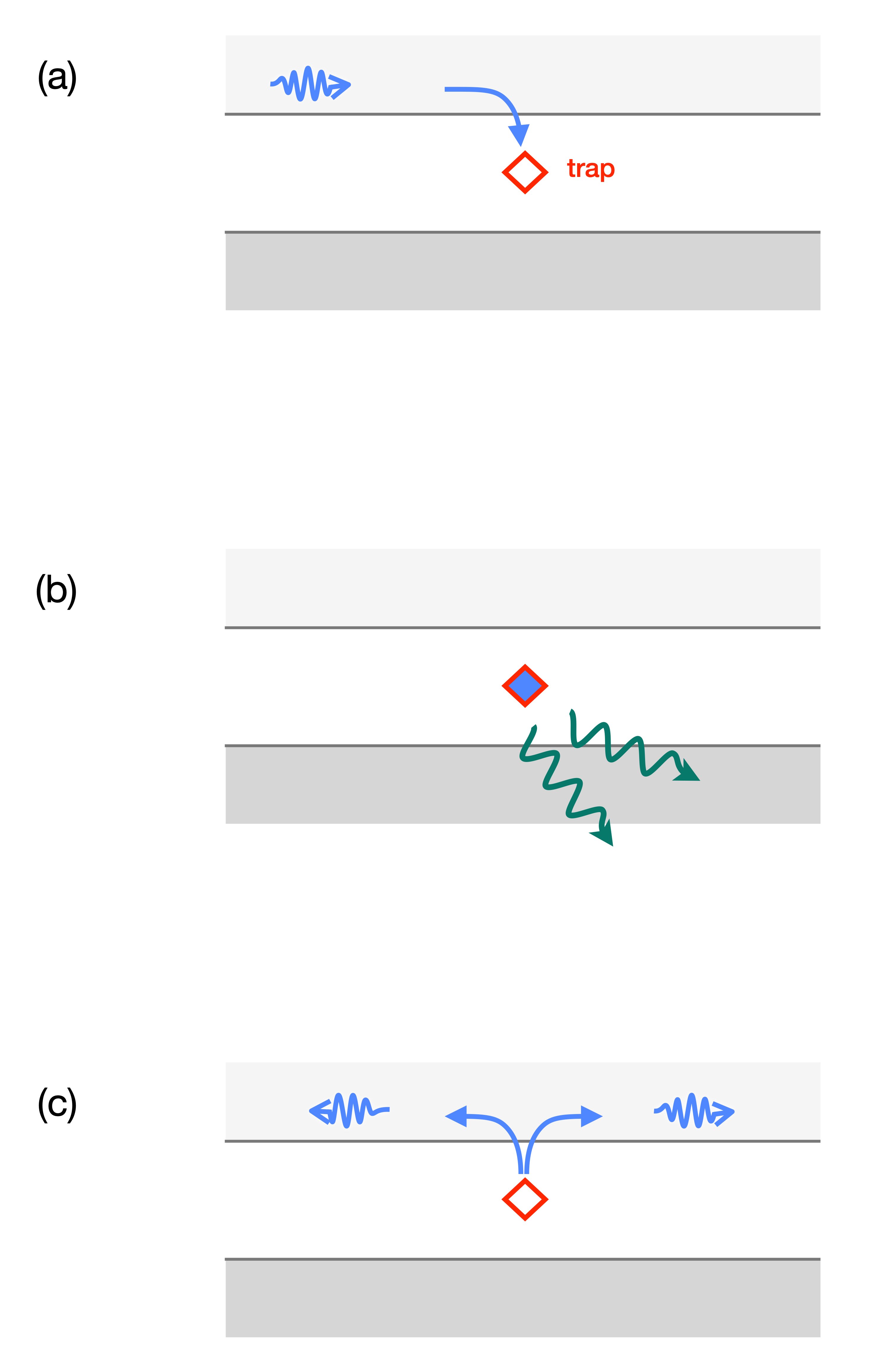}
\caption[width=0.6] {Band diagrams illustrating the interaction of an electron (blue wavepacket) with a deep track in a semiconductor. An electron arriving from the left is caught by the trap (a), thereby emitting one or several phonons (green arrows), which disappear in the thermal bath (b). The electron is localized at the trap until it is released by a phonon fluctuation. It then propagates either to the left or to the right or even coherently in both directions~(c).}
\centering
\end{figure}

Now we can pipeline this non-unitary process with a unitary evolution designed to feed the trap site with incoming wave packets such that a noteworthy state evolution results.

\subsection{Unitary evolution}

In the nonrelativistic regime, the unitary evolution of a quantum state is described precisely by Schr\"odinger's equation. We are free to choose the device architecture in which the quantum state evolves. For example, we consider a quantum ring that is biased with a magnetic field $H$, which penetrates the ring hole. The lengths of its two arms are chosen to differ slightly, $\Delta l = l_1 - l_2 \ne 0$ (Fig.~4). If the magnetic field has a well-defined value $H^*$, the unitary evolution of electron wave packets is as follows~\cite{Bredol2019, MannhartBraak2018}.

\begin{figure}[t]
\centering
\includegraphics[width=0.4\textwidth]{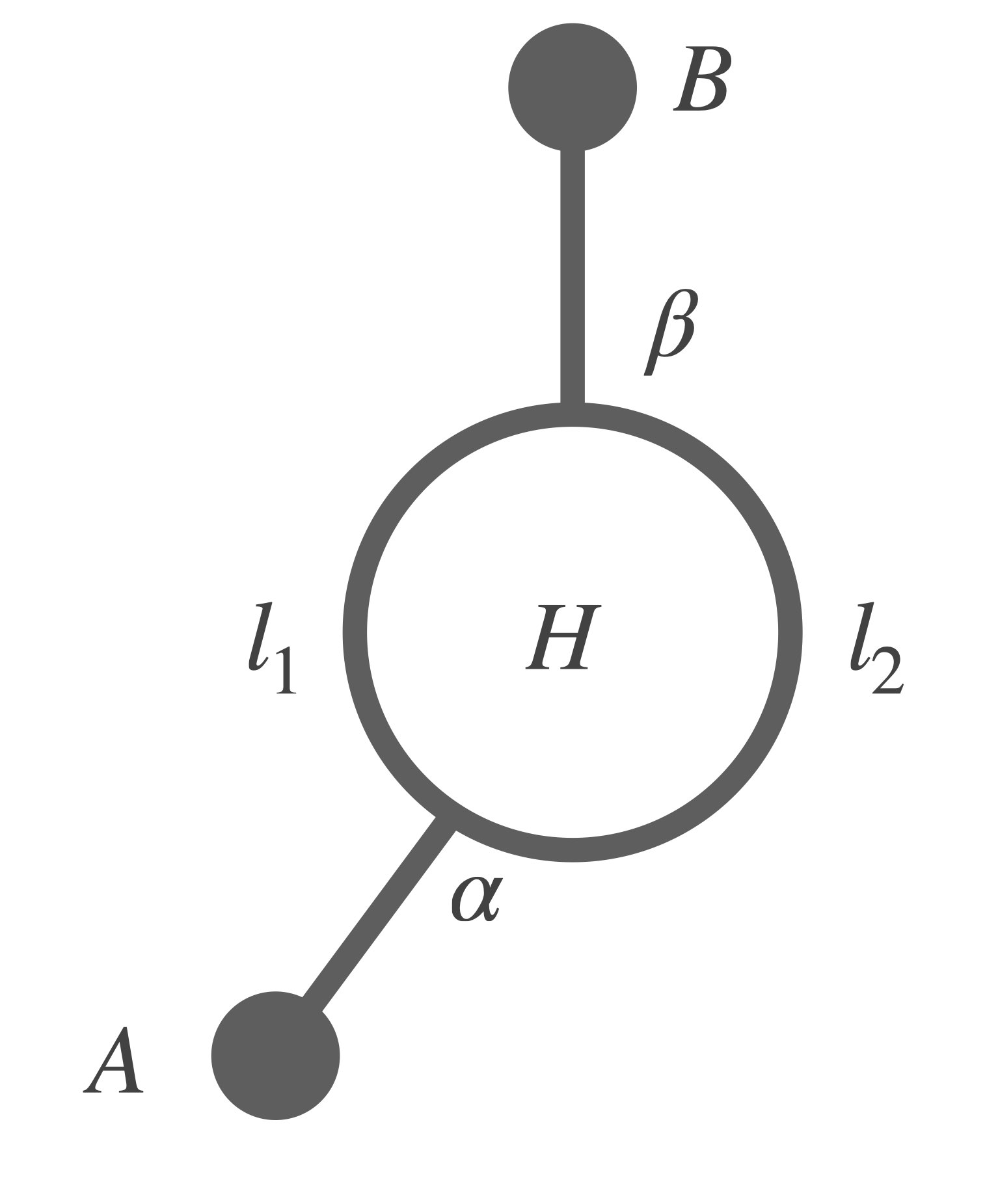}
\caption[width=0.7] {Sketch of an asymmetric quantum ring that connects  two terminals A and B via the junctions $\alpha$ and $\beta$. The two arms of the ring that connect $\alpha$ and $\beta$ have lengths $l_1$ and $l_2$. The ring is penetrated by a magnetic field~$H$.}
\centering
\end{figure}

\textit{Electron transport from terminal A to terminal B: } 
Electrons arriving from terminal A are split at the junction $\alpha$ (Fig.~4) into two wave packets that then enter the ring. Owing to $ H^*$ and $\Delta l$, these two wave packets acquire on their way to junction $\beta$ a mutual phase difference of $\Delta \varphi = 2n \pi$, 
where $n$ is a natural number. When the two wave packets meet at the top of the ring at junction $\beta$, they therefore interfere constructively, and the resulting wave packet leaves the ring and heads to terminal B. The length of the electron trajectory in the ring is $l_{\textrm{A} \to \textrm{B}} \approx \pi R$, where $R$ is the ring's radius. The corresponding electron transit time is denoted $ \tau_{\textrm{A} \to \textrm{B}}$.

\textit{Electron transport from terminal B to terminal A: }  
Electrons arriving from terminal B split at junction $\beta$ and then move to $\alpha$. These two wave packets acquire a mutual phase difference of $\Delta \varphi = (2m+1) \pi$, where $m$ is a natural number, so that they interfere destructively at $\alpha$. They re-arrive at $\beta$ with $\Delta \varphi = (2n + 2m + 1) \pi$. As their phase difference is still an odd multiple of $\pi$, destructive interference hits again. The wave packets are reflected back to $\alpha$, where they interfere with a mutual phase difference of $\Delta \varphi = (2n + 4m + 2) \pi$, so that they now leave the ring for terminal A. The length of their trajectory is $l_{\textrm{B} \to \textrm{A}} \approx 3 \pi R$ and their transit time is $\tau_{\textrm{B} \to \textrm{A}} = 3\tau_{\textrm{A} \to \textrm{B}}$, i.e.\ triple those of the $\textrm{A} \to \textrm{B}$ direction.

The asymmetric quantum ring is therefore characterized by a nonreciprocity of the transmission time. This temporal nonreciprocity varies with $H$ in an oscillatory manner between $1/3$ and $3$. The transmission probability is reciprocal as demanded by unitarity (see, for example, \cite{Datta}) at $H=H^*$ equaling $\sim \! 1$ for both directions. The same behavior is induced by symmetric quantum rings pattered into interfaces that are subject to strong Rashba coupling and a Zeeman field \cite{Mannhart2018}. Indeed, as shown in \cite{Bredol2019,Bredol2020}, nonreciprocities of particle transmission or reflection times are a characteristic feature of noncentrosymmetric unitary conductors. The arrow-shaped device in Fig.~5 (top) provides an example. As noted many years ago, related nonreciprocal phenomena also exist in optics, see for example Rayleigh's paradox~\cite{Rayleigh1885,Rayleigh1901}. 

\subsection{Unitary and non-unitary evolutions combined in an asymmetric quantum ring}

By designing the quantum ring to be asymmetric and by applying a magnetic field, we utilize the unitary state evolution to acquire a nonreciprocal transmission time and travel path length. Accordingly, an electron passes any point of the ring three times or only once, depending on whether the electron travels from B to A or from A to B. The combination of this tailored unitary state evolution with the non-unitary evolution induced by phase breaking at the trapping site yields nonreciprocal transmission probabilities in addition to the nonreciprocal transmission times and trajectory lengths.

Referring back to Figure 2, let us recall an asymmetric quantum ring with an embedded electron trapping center. We consider the case that the decoherence time, or the mean inelastic scattering time, equals $\tau_\textrm{in} = 2 \tau_{\textrm{A} \to \textrm{B}}$, which is twice the time an $\textrm{A} \to \textrm{B}$ electron takes to cross the ring unitarily. As $\tau_{\textrm{B} \to \textrm{A}} = 3 \tau_{\textrm{A} \to \textrm{B}}$, the trapping site will seize three times more $\textrm{B} \to \textrm{A}$ than $\textrm{A} \to \textrm{B}$ electrons. As the trap re-emits an electron without directional preference, it creates an imbalance in the transmission probability by sending back a disproportionately large number of electrons arriving from B. As a result, these rings let electrons pass preferably in one direction, namely from A to~B.

A nonreciprocal transmission probability for single electrons passing a two-terminal device is beyond Onsager's reciprocal relation. Furthermore, the ring’s properties challenge the second law of thermodynamics. In view of these possibly far-reaching implications, it is desirable to model the behavior of the devices quantitatively. We have therefore analyzed the electron transport across such quantum rings by independent numerical simulations~\cite{Bredol2019}.

In these simulations, at the start of the time evolution, the electron wave functions are described by Gaussian wave packets. Their unitary time evolutions are calculated by numerically solving Schr\"odinger's equation on the ring geometry employing a tight-binding model and exact diagonalization. The stochastic trapping events are determined in a Monte Carlo algorithm based on Born's rule and on the projection of the wave function onto localized states at the trap site. The electrons are then released again, forming new wave packets. If no trapping was found to take place, i.e., if a null-measurement occurred, the complementary projection is applied, which erases the wave function at the trap sites. The evolution of the wave function after the occurrence or non-occurrence of a trapping event is then calculated again in accordance with  Schr\"odinger's equation until the next trapping event occurs and the above procedure repeats. This loop is quit when the electron is found to have reached a contact. The Monte Carlo character of this calculation provides a stochastic transmission probability. A Monte Carlo algorithm is used to find the expectation values of the resulting stochastic transmission probabilities.

Figure 5 shows the calculated difference of the mean transmission probabilities $P_{\textrm{A} \to \textrm{B}} - P_{\textrm{B} \to \textrm{A}}$ of electrons traveling in the $\textrm{A} \to \textrm{B}$ and $\textrm{B} \to \textrm{A}$ directions in a device as shown in the panel at the top. As the shape of such devices  is asymmetric with respect to the direction of the current flow, even without an applied magnetic field, the transport properties of these devices are comparable to the ones of asymmetric quantum rings that are magnetic-field biased \cite{Bredol2019}. The transmission probabilities were calculated as a function of the mean inelastic scattering time, which was varied as a free parameter. The graph shows the probabilities as a function of the resulting average inelastic scattering events per electron transmission. If the number of scattering events is insignificant, this corresponds to the left-hand side of the diagram where the evolution is almost exclusively unitary and inelastic scattering is too rare for the non-unitary evolution to be relevant. In agreement with this unitary evolution, the transport probabilities are reciprocal, $P_{\textrm{A} \to \textrm{B}} - P_{\textrm{B} \to \textrm{A}} \approx 0$. This behavior is consistent with Onsager's reciprocal relation. 

\begin{figure}[t]
\centering
\includegraphics[width=1.0\textwidth]{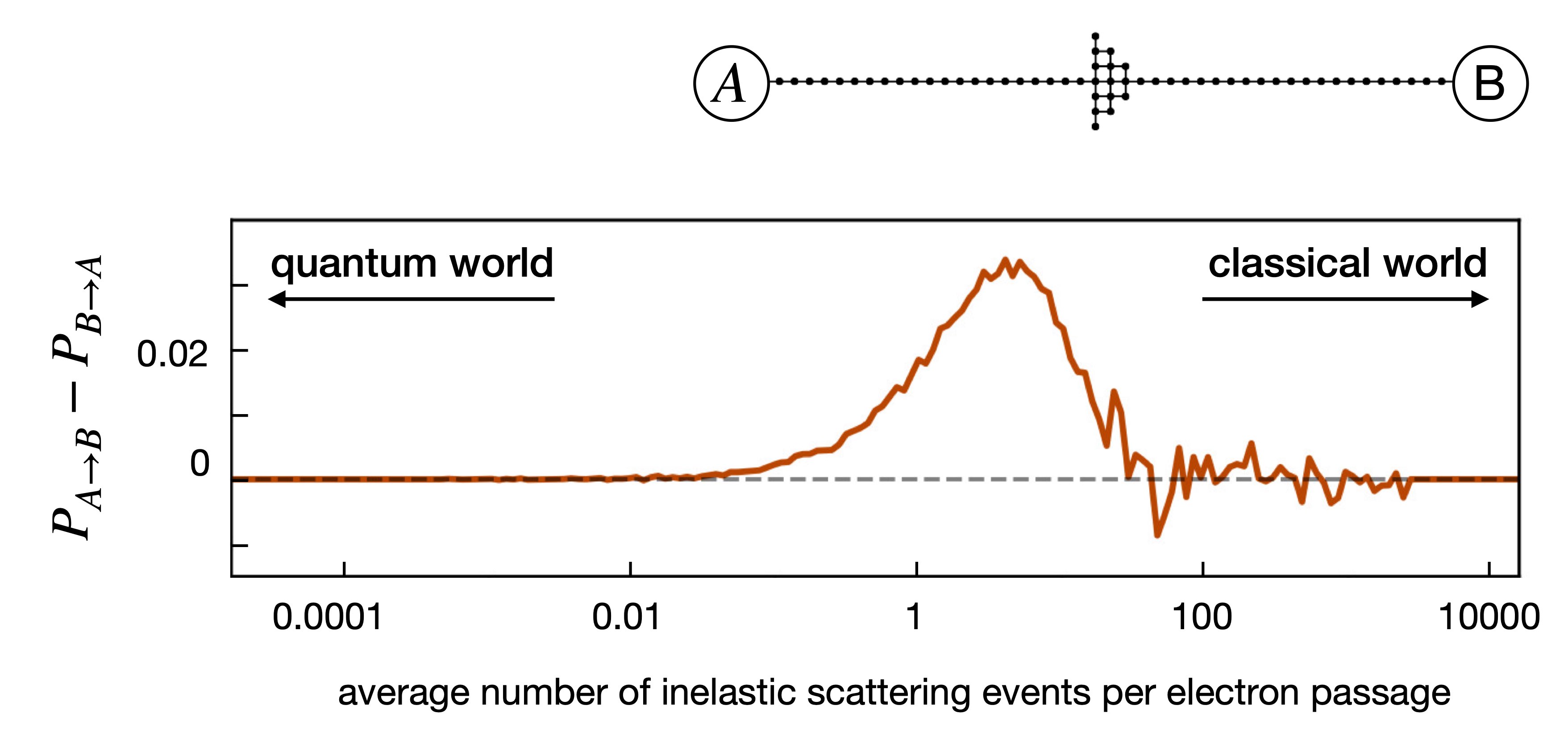}
\caption {Difference of the transmission probabilities $P_{\textrm{A} \to \textrm{B}}$ and $P_{\textrm{B} \to \textrm{A}}$ calculated as a function of the the average number of inelastic scattering events per electron passage, which scales with the inverse phase breaking length. The device is sketched at the top with  dots presenting the sites used in the tight-binding model. The sorting function of the device is shown on the right. Signiﬁcant sorting is achieved for $0.05-50$ inelastic scattering events when the transport deviates from reciprocity (after~\cite{Bredol2019})}
\end{figure}

If the average number of scattering events reaches the order of one per transmission, this is the case where the decoherence time is comparable to the characteristic device transmission time, and $ P_{\textrm{A} \to \textrm{B}} - P_{\textrm{B} \to \textrm{A}}$ achieves a finite value. It reaches a maximum of $\approx \! 3 \%$ at an average on the order of two events per transmission. 

For a much larger number of scattering events, i.e.\ for very short decoherence times, the unitary phase evolution is overwhelmed by the non-unitary processes. With interference becoming irrelevant and the electrons undergoing scattering events that are so numerous, they statistically average, and the electron transport acquires an increasingly classical character. Consistent with the time-reversal symmetry of the classical transport and Onsager's relation, the transport is then reciprocal again. Sorting occurs neither in the classical, nor in the unitary quantum regime: $P_{\textrm{A} \to \textrm{B}} - P_{\textrm{B} \to \textrm{A}} \approx 0$.

Sorting takes place in the well-defined transition regime of $\sim \! 0.05-50$ scattering events per transmission, and only there. In linear response, the transmission probability of the two-terminal device is nonreciprocal in this regime. This transport does not follow Onsager's reciprocal relation \cite{Bredol2019}. Indeed, it does not have to, because Onsager's relation is built on the assumption of time reversibility of the microscopic behavior. In our case, this reversibility is given only in the unitary quantum regime and in the classical realm. In both regimes, the transport of the ring does indeed follow Onsager's relation.

Focusing on the transition regime, we point out that the nonreciprocal conductance of such a ring differs greatly from the nonreciprocal characteristics of standard diodes, say $pn$-diodes. These diodes consist of two materials with different chemical potentials for electrons. In thermal equilibrium and without applied bias voltage, the electrochemical potential is spatially constant across the diode, which is achieved by a variation of the electron density, which lifts the electrostatic potential of the diode's $p$-side with respect to its $n$-side. In thermal equilibrium, the resulting drift current cancels the diffusion current of the electrons. Nonreciprocal conductance occurs exclusively at finite bias voltages. In linear response, the conductance is linear. In contrast, the non-unitary, asymmetric quantum ring requires only a single material. It is the shape of the ring and the phase-breaking scattering that induce a nonreciprocal conductance, also in linear response.

The nonreciprocal transmission probability for individual electrons, depending on whether they travel from A to B or from B to A, which we have derived from the two axioms of quantum mechanics, corresponds to the behavior of a demon as proposed by Maxwell \cite{Maxwell,Leff2003,Capek,Lutz} and therefore challenges the second law of thermodynamics. As the microscopic dynamics of the rings breaks time-reversal symmetry, it is understandable that these non-unitary devices do not obey Onsager's reciprocal relation in terms of linear response either. Is there a comparable rationale why the second law might not apply to the rings? Well, it is accepted that a functioning demon, as introduced by Maxwell, would break the second law. The explanation that is currently accepted by most members of the community as to why such a demon cannot exist seems compelling: To do its job in a cyclic process, the demon has to  open and close a gateway actively, and for this it has to acquire, process, store and therefore also discard information about the particles \cite{Lutz,Penrose,Bennett}. The information storage must be stable against thermal fluctuations. Hence, the energy well stabilizing a stored bit has to be large compared to $kT$. After the door has been actuated, the stored information must be erased, which is done by resetting a memory to a defined value. If the data storage cell were not cleared, the demon would pile up information, which would be inconsistent with a cyclic process. As Landauer's erasure theorem \cite{Landauer} proves that the reset of a memory requires dissipation per bit of at least $W = k_\textrm{B} T \ln  2$ to the thermal bath, the demon cannot violate the second law~\cite{Lutz}. 

Much as Onsager's relation rests on the assumption that processes are time-reversal symmetric, this erasure argument uses the implicit assumption that an intelligence-driven action of opening and closing a door is required. It is the control of this actuation that involves the processing, storage and erasure of information. However, the validity of this assumption is not obvious~\cite{Leff2003, Norton2004}. 

In the non-unitary quantum rings, actuation or control of a door or gate is not required. At any time, the path through the ring simultaneously has a higher transmission for electrons coming from the left than for electrons from the right. When a defect traps an electron in a thermally equilibrated ring, the information contained in the phase and momentum of the incoming wave packet does not need to be stored in a memory cell that is stable against thermal activation. The energy of the electron and therefore the energy required to alter information is always within the thermal energy range ($\sim \! kT$) of the trap site's energy. The information contained in the incoming wave packet cannot pile up, because it is not written into a stable storage cell, but disappears with the generated phonons into the thermal bath. During the trapping process, the quantum state of the electron is projected onto the ground state of the defect. The von~Neumann projection at the thermally anchored defect thereby generates a new and well-defined state for all original electron travel directions. For this, a Landauer-type reset of a data storage cell, which would require an energy input of $W = kT \ln 2$ and an associated dissipation, is not needed. When a thermal fluctuation releases the electron again from the trap site, a new information-carrying quantum state is created, again without the need for an external energy input. Therefore, the Landauer erasure argument against the existence of second-law-violating Maxwell demons does not pertain to such non-unitary processes. 

Indeed, the request that the demon be intelligent goes back to von~Smoluchowski \cite{Smoluchowski1912}, who showed that non-intelligent demons based on classical mechanics cannot function. At that time, he could not consider non-unitary devices, and his arguments do not rule out demons such as our devices that are neither intelligent nor based on classical mechanics.

It is remarkable that sorting is achieved by having a source of random noise (the trap site coupled to the thermal bath). This noise source  breaks the balance between the ${\textrm{A} \to \textrm{B}}$ and ${\textrm{B} \to \textrm{A}}$ channels, which are actuated by Johnson noise. A noise source disturbing the balance of two noisy channels therefore creates an imbalance,  a net electron current that flows from ${\textrm{A} \to \textrm{B}}$. This astounding behavior is reminiscent of a curious phenomenon known in game theory as Parrondo's paradox: 
A random switching between two games of chance, both of which are either unbiased or even skewed to lose, yields on average a winning outcome \cite{Harmer19991,Harmer19992,Astumian,Key2006}. 
Interestingly, exactly this non-intuitive behavior has been used to elucidate the operation of molecular motors~\cite{Astumian}.

For this summer school, it is of particular interest that the principle of combining unitary and non-unitary quantum state evolutions to achieve phenomena that violate the existing rules and laws does not apply exclusively to the non-unitary quantum rings presented but covers a much wider space. To stay first with two-terminal electronic devices, we note that the rings can also function as devices that transport particles in one direction more as waves, and in the reverse direction as particles. They also provide for devices that transport particles phase-coherently in forward direction, and incoherently in reverse direction, or, in other words, devices that feature quantum transport in one direction, and more classical transport in the other \cite{Mannhart2019}. The basic principles require neither rings nor are they specific to electrons \cite{MannhartBraak2018}. As shown in \cite{Braak2020}, photonic devices coupled to chiral waveguides behave inconsistently with the second law if the photon evolution comprises unitary and non-unitary steps. 

The principle to apply unitary and non-unitary evolutions of quantum states to attain novel functions far exceeds transport in electronic or photonic devices. It is worthwhile to consider, for example, whether this principle can also be applied to alter the balance of chemical reactions. Coherent state evolutions and interference are relevant for many chemical reactions; some reactions even unfold as coherent superpositions of distinct pathways \cite{Dai2003, Aoiz2018}. The principle is also applicable to realize materials with novel properties, as we will show now. 

\section{Non-unitary Quantum Materials}

The possibly valuable functions of non-unitary quantum materials prompt the question of whether nature has already discovered ordering phenomena induced by mixing unitary and non-unitary processes, for example in order to synthesize complex molecules or to harvest thermal energy. Whether nature achieves ordering phenomena or other novel functions by such processes, we do not know. It is known, however, that evolution has generated many biomolecules that comprise unitary and non-unitary processes to tune conductances, as we will illustrate with two examples. The first example refers to a large set of electrically conducting biomolecules. As discussed in Ref.~\cite{Vattay2015}, numerous conducting biomolecules are tuned to the critical point between the metal-insulator transition separating the Anderson-insulator phase from the conducting disordered metal, which is highly improbable to happen by random chance. At this transition, the electron  transport in the molecules is subject to both coherent propagation and decoherence caused by the environment \cite{Vattay2015}. Our second example is the Fenna--Matthews--Olson complex, a light-harvesting complex in green sulfur bacteria. For this complex, the balanced coupling of coherent and incoherent processes has been found to maximize the excitonic energy transfer through the complex, which is relevant for the evolutionary fitness of the bacteria~\cite{Pelzer2014}.

Materials can also be artificially designed and metamaterials can be devised to have properties shaped by the non-unitary and unitary evolution of quantum states \cite{Mannhart2019}, see for example Figure~6. This 2D material consists of ring-shaped molecules that break inversion symmetry, where each molecule acts as an asymmetric, non-unitary quantum ring. The molecules are incoherently coupled with their neighbors by phase-breaking contacts. Such a material will not follow Onsager's relation, but instead will show macroscopic nonreciprocal transport in linear response. Of course, such 2D planes may be stacked to form 3D crystals or heterostructures with even more complex properties at their interfaces.

\begin{figure}[t]
\centering
\includegraphics[width=0.8\textwidth]{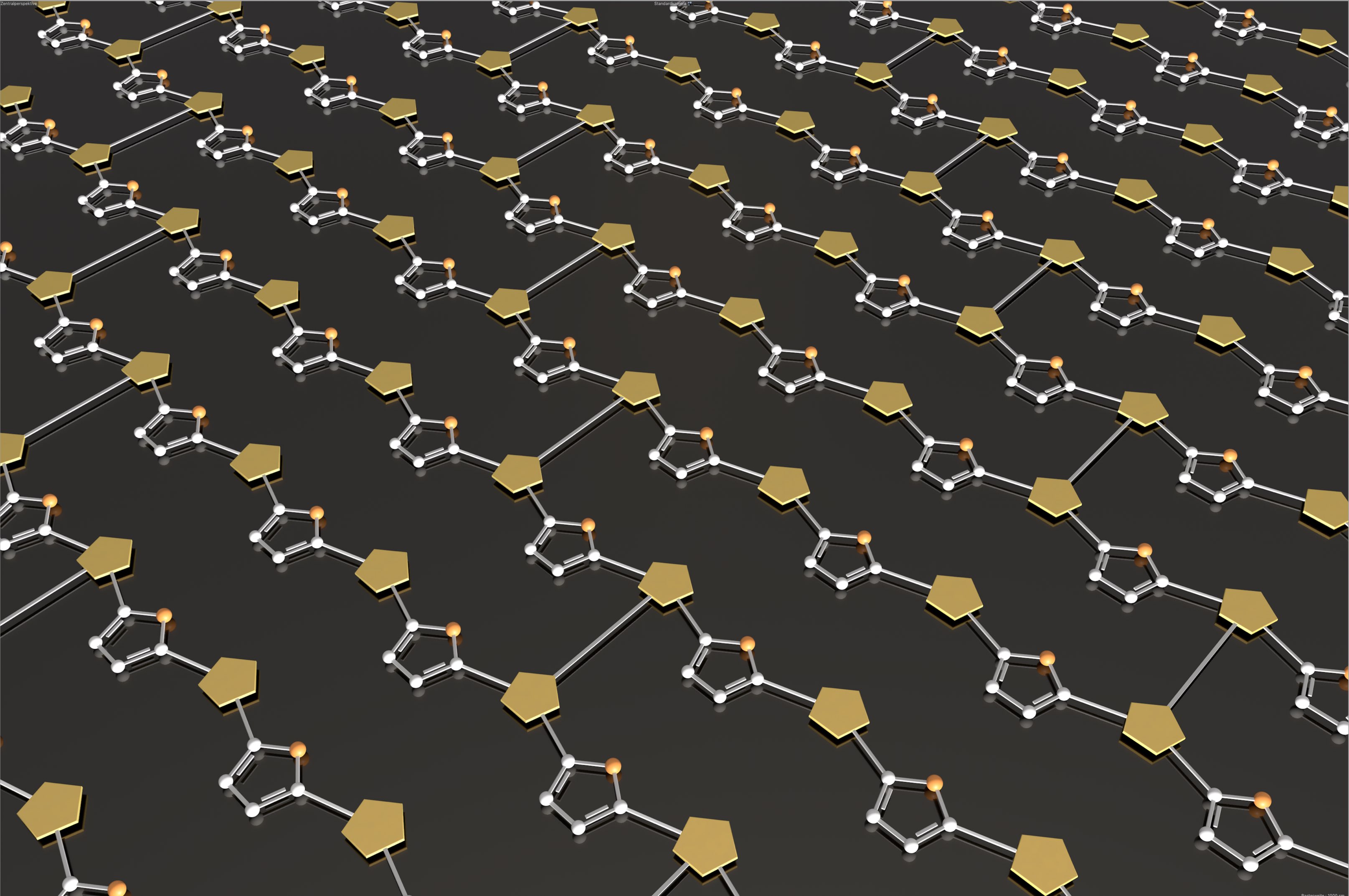}
\caption {Schematic illustration showing the principle of implementing the function of noncentrosymmetric and non-unitary quantum devices into macroscopically large objects. The sketch renders a macroscopic array of conducting molecules forming asymmetric loops. These molecules connect to ports (gold) with incoherent electron systems (from~\cite{Bredol2019}).}
\end{figure}

\section{Conclusions}

From the above, it follows that the transition regime between the quantum and the classical world offers  possibilities for devices with unheard-of functions. State projections 
provide a valuable asset for device applications, and they open new horizons for electronics and quantum materials. However, their potential is overlooked if they are considered but a nuisance because they cause  unitary quantum states to decohere. 

Non-unitary electronics and non-unitary quantum materials are an exemption to Onsager's reciprocal relation and the laws and rules built upon it. The devices also challenge the second law of thermodynamics. It does not follow from our Gedankenexperiment that the second law of thermodynamics does not apply to the transition regime. Instead, the argument presented here shows that the axioms of quantum physics do not agree with the second law of thermodynamics. Which of the two is correct? To us, the arguments in favor of quantum physics seem strong, but the ultimate answer will be provided by experiments, and for these we propose photonic systems or electronic devices as described here.

The transition regime---the edge of the quantum world---is a place that may harbor further exemptions from laws that are well accepted in the classical and unitary quantum domains and therefore thought to be universally valid. It is a place worthwhile exploring for new devices and materials.

\section*{Acknowledgments}

The authors gratefully acknowledge close and beneficial interactions with Daniel Braak,  helpful discussions with numerous other colleagues, and outstanding editorial support by Lilli Pavka. The numerical calculations were performed using the Kwant \cite{Groth} Python package.

\end{document}